\documentclass[preprint,12pt]{elsarticle} 
\usepackage{graphicx} 
\usepackage{amssymb} 
\usepackage{amsmath} 
\usepackage{amsfonts} 
\usepackage[dvipsnames]{xcolor} 
\def\lamb#1#2{$^{#1}_{\Lambda}${#2}}

\newcommand{\be}{\begin{equation}} 
\newcommand{\ee}{\end{equation}}

\newcommand{\half}{\frac{1}{2}} 
\journal{Physics Letters B} 

\begin{document}

\begin{frontmatter}

\title{Towards resolving the \lamb{3}{H} lifetime puzzle} 

\author[a]{A.~Gal\corref{cor1}~} 
\author[b]{~H.~Garcilazo} 
\address[a]{Racah Institute of Physics, The Hebrew University, 91904 
Jerusalem, Israel}
\cortext[cor1]{corresponding author: Avraham Gal, avragal@savion.huji.ac.il}  
\address[b]{Escuela Superior de F\' \i sica y Matem\'aticas \\ 
Instituto Polit\'ecnico Nacional, Edificio 9, 07738 M\'exico D.F., Mexico} 

\date{\today} 

\begin{abstract} 

Recent \lamb{3}{H} lifetime measurements in relativistic heavy ion collision 
experiments have yielded values shorter by (30$\pm$8)\% than the free 
$\Lambda$ lifetime $\tau_\Lambda$, thereby questioning the naive expectation 
that $\tau$(\lamb{3}{H})$\,\approx\,$$\tau_\Lambda$ for a weakly bound 
$\Lambda$ hyperon. Here we apply the closure approximation introduced 
by Dalitz and coworkers to evaluate the \lamb{3}{H} lifetime, using 
\lamb{3}{H} wavefunctions generated by solving three-body Faddeev 
equations. Our result, disregarding pion final-state interaction (FSI), 
is $\tau$(\lamb{3}{H})=(0.90$\pm$0.01)$\tau_\Lambda$. In contrast to previous 
works, pion FSI is found attractive, reducing further $\tau$(\lamb{3}{H}) 
to $\tau$(\lamb{3}{H})=(0.81$\pm$0.02)$\tau_\Lambda$. We also evaluate for 
the first time $\tau$(\lamb{3}{n}), finding it considerably longer than 
$\tau_\Lambda$, contrary to the shorter lifetime values suggested by the 
GSI HypHI experiment for this controversial hypernucleus. 

\end{abstract} 


\end{frontmatter}

\section{Introduction} 
\label{sec:intro} 

\begin{figure}[htb]
\begin{center}
\includegraphics[width=0.92\textwidth]{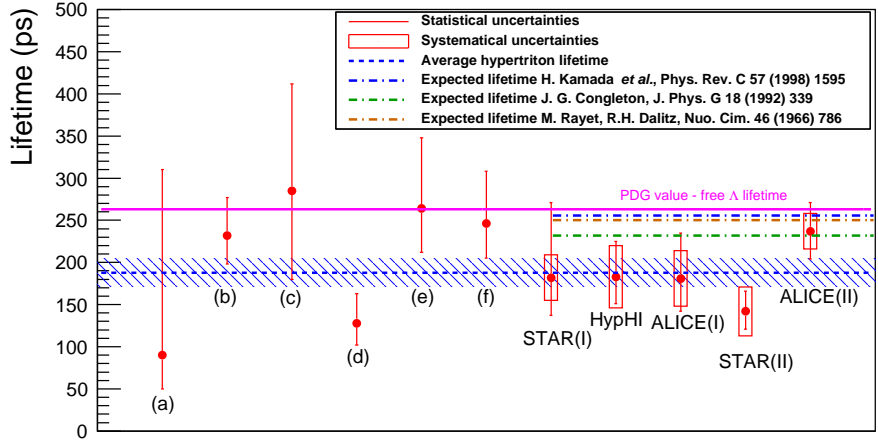}
\caption{Measured \lamb{3}{H} lifetime values in chronological 
order, with (a)--(f) from emulsion and bubble-chamber 
measurements~\cite{PS64,Keyes68,PS69,Bohm70,Keyes70,KSWB73}, and from 
recent relativistic heavy ion experiments: STAR(I)~\cite{STAR10}, 
HypHI~\cite{HypHI13}, ALICE(I)~\cite{ALICE16}, STAR(II)~\cite{STAR18}, 
ALICE(II)~\cite{ALICEprel}, see text. We thank Benjamin D\"{o}nigus 
for providing this figure~\cite{BMD19}.} 
\label{fig:L3H}
\end{center}
\end{figure}

\lamb{3}{H}, a $pn\Lambda$ state with spin-parity $J^P=\half^+$ and isospin 
$I=0$ in which the $\Lambda$ hyperon is bound to a deuteron core by merely 
$B_{\Lambda}$(\lamb{3}{H})=0.13$\pm$0.05~MeV, presents in the absence of 
two-body $\Lambda N$ bound states the lightest bound and one of the most 
fundamental $\Lambda$ hypernuclear systems~\cite{GHM16}. Its spin-parity 
$\half^+$ assignment follows from the measured branching ratio of the two-body 
decay \lamb{3}{H}~$\to{^3{\rm He}}+\pi^-$ induced by the free $\Lambda$ 
weak decay $\Lambda\to p+\pi^-$~\cite{Block64}. There is no experimental 
indication, nor theoretical compelling reason, for a bound $J^P={\frac{3}{2}
}^+$ spin-flip excited state, and there is even less of a good reason to 
assume that an excited $I=1$ state lies below the $pn\Lambda$ threshold. 

Given the loose binding of the $\Lambda$ hyperon in \lamb{3}{H} it is natural 
to expect, perhaps naively, that nuclear medium effects modify little the 
free $\Lambda$ lifetime in such a diffuse environment. An updated compilation 
of measured \lamb{3}{H} lifetime values is presented in Fig.~\ref{fig:L3H}. 
Note the world-average value (dashed) which is shorter by about 30\% than the 
free $\Lambda$ lifetime $\tau_\Lambda=263\pm 2$~ps (solid). In sharp contrast 
with the large scatter of bubble chamber and nuclear emulsion measurements 
from the 1960s and 1970s, the recent measurements of $\tau$(\lamb{3}{H}) 
in relativistic heavy ion experiments marked in the figure give values 
persistently shorter by (30$\pm$8)\% than $\tau_\Lambda$~\cite{BMD19}. 
Also shown are \lamb{3}{H} lifetime values from three calculations 
(dot-dashed) that pass our judgement, two of which~\cite{RD66,Fad98} 
using fully three-body \lamb{3}{H} wavefunctions claim \lamb{3}{H} lifetimes 
shorter than $\tau_\Lambda$ by only (4$\pm$1)\%. The third one \cite{Con92}, 
using a $\Lambda d$ cluster wavefunction, obtained a \lamb{3}{H} lifetime 
shorter than $\tau_\Lambda$ by as much as 13\%. These cited results include 
a small nonmesonic decay rate contribution of 1.7\%~\cite{Golak97}. Among 
calculations that claim much shorter \lamb{3}{H} lifetimes, we were unable 
to reproduce the results of Ref.~\cite{BP69}, nor to make sense out of 
a \lamb{3}{H} decay rate calculation based on a nonmesonic $\Lambda N\to NN$ 
weak interaction hamiltonian~\cite{MH79}. We comment briefly on the 
calculations in Refs.~\cite{RD66,Fad98}: 

(i)~~Rayet and Dalitz (RD)~\cite{RD66}, using a closure approximation to sum 
over the final nuclear states reached in the \lamb{3}{H} weak decay, reduced 
the \lamb{3}{H} lifetime calculation to the evaluation of a \lamb{3}{H} 
exchange matrix element defined in Sect.~\ref{sec:Tot} below. Variational 
\lamb{3}{H} wavefunctions of the form $f(r_{\Lambda p})f(r_{\Lambda n})
g(r_{pn})$ were used, accounting for both short-range and long-range 
correlations in the diffuse \lamb{3}{H}. With a suitable choice of the closure 
energy, and including a questionable 1.3\% repulsive pion FSI decay-rate 
contribution (see below), they obtained $\tau$(\lamb{3}{H})$\,\approx 
0.95\,\tau_\Lambda$. However, the RD decay rate expressions miss a recoil 
phase-space factor which, if not omitted inadvertently in print, would 
bring down their calculated \lamb{3}{H} lifetime to 85\% of $\tau_\Lambda$ 
(see footnote b to table 4 in Ref.~\cite{KSWB73} to support this scenario). 

(ii)~~Kamada et al.~\cite{Fad98} in a genuinely {\it ab-initio} calculation 
used a \lamb{3}{H} wavefunction obtained by solving three-body Faddeev 
equations with $NN$ and $YN$ Nijmegen soft-core potentials to evaluate all 
three $\pi^-$ decay channels: $^3$He + $\pi^-$, $d+p+\pi^-$ and $p+p+n+\pi^-$. 
The $\pi^0$ decay channels were related by the $\Delta I =\half$ rule in a 
ratio 1:2 to the corresponding $\pi^-$ channels. Their calculated \lamb{3}{H} 
lifetime is 256~ps: shorter by 3\% than the measured value of $\tau_\Lambda$, 
but shorter by 6\% than their calculated value of 272~ps for $\tau_\Lambda$. 
Hence, we refer to their result as $\tau$(\lamb{3}{H})$\,\approx 0.94\,
\tau_\Lambda$. 

In this Letter we study pion FSI which in accord with low-energy pion-nucleus 
phenomenology~\cite{BFG97,FG07} is generally considered repulsive, thereby 
increasing rather than decreasing $\tau$(\lamb{3}{H}). However, exceptionally 
for \lamb{3}{H}, pion FSI is attractive and potentially capable of resolving 
much of the $\tau$(\lamb{3}{H}) puzzle. A fully microscopic inclusion of 
pion FSI requires a four-body final-state model, a formidable project that 
still needs to be done. Instead, we study here $\tau$(\lamb{3}{H}) within 
a closure-approximation calculation in which the associated exchange matrix 
element is evaluated with wavefunctions obtained by solving \lamb{3}{H} 
three-body Faddeev equations. Disregarding pion FSI, our result $\tau
$(\lamb{3}{H})$\,\approx 0.90\,\tau_\Lambda$ differs by a few percent from 
that of the microscopic Faddeev calculation by Kamada et al.~\cite{Fad98}. 
Introducing pion FSI in terms of pion distorted scattering waves results in 
$\tau$(\lamb{3}{H})=(0.81$\pm$0.02)$\tau_\Lambda$, that is (213$\pm$5)~ps, in 
the right direction towards resolving much of the $\tau$(\lamb{3}{H}) puzzle. 

Finally, as a by-product of studying $\tau$(\lamb{3}{H}), we estimate for the 
first time the lifetime of \lamb{3}{n} assuming that it is bound. The particle 
stability of \lamb{3}{n} was conjectured by the GSI HypHI Collaboration having 
observed a $^3$H+$\pi^-$ decay track~\cite{Rap13}, but is unanimously opposed 
by recent theoretical works~\cite{GV14,Hiyama14,GG14}. Our estimate suggests 
a value of $\tau$(\lamb{3}{n}) considerably longer than $\tau_\Lambda$, in 
strong disagreement with the shorter lifetime reported in Ref.~\cite{Rap13}.

\section{Total decay rate expressions for \lamb{3}{H} and \lamb{3}{n}} 
\label{sec:Tot} 

The $\Lambda$ weak decay rate considered here, $\Gamma_{\Lambda}\approx
\Gamma_{\Lambda}^{\pi^-}+\Gamma_{\Lambda}^{\pi^0}$, accounts for the 
mesonic decay channels $p\pi^-$ (63.9\%) and $n\pi^0$ (35.8\%). Each of 
these partial rates consists of a parity-violating $s$-wave term (88.3\%) 
and a parity-conserving $p$-wave term (11.7\%), summing up to 
\begin{equation} 
\Gamma_\Lambda(q)=\frac{q}{1+\omega_{\pi}(q)/E_N(q)}
(|s_{\pi}|^2+|p_{\pi}|^2\frac{q^2}{q_{\Lambda}^2}),~~~~~ 
\left|\frac{p_{\pi}}{s_{\pi}}\right|^2 \approx 0.132, 
\label{eq:GammaLam} 
\end{equation} 
where $\Gamma_\Lambda$ is normalized to $|s_{\pi}|^2+|p_{\pi}|^2=1$, 
$\omega_{\pi}(q)$ and $E_N(q)$ are center-of-mass (cm) energies of the decay 
pion and the recoil nucleon, respectively, and $q\to q_{\Lambda}\approx 102
$~MeV/c in the free-space $\Lambda\to N\pi$ weak decay. The $\approx$2:1 ratio 
of $\pi^-$:$\pi^0$ decay rates, the so called $\Delta I = \frac{1}{2}$ rule in 
nonleptonic weak decays, assigns the final $\pi N$ system to a well-defined 
$I=\frac{1}{2}$ isospin state.

\subsection{\lamb{3}{H}} 
\label{subsec:L3H} 

For \lamb{3}{H} ground state (g.s.) weak decay, approximating the outgoing 
pion momentum by a mean value $\bar q$ and using closure in the evaluation 
of the summed mesonic decay rate, one obtains~\cite{RD66} 
\begin{equation} 
\Gamma_{_\Lambda^3{\rm H}}^{J=1/2}=\frac{\bar q}{1+\omega_{\pi}(\bar q)
/E_{3N}(\bar q)}[|s_{\pi}|^2(1+\frac{1}{2}\eta(\bar q))+|p_{\pi}|^2(
\frac{\bar q}{q_{\Lambda}})^2(1-\frac{5}{6}\eta(\bar q))].  
\label{eq:L3Hg.s.} 
\end{equation}  
In this equation we have omitted terms of order 0.5\% of $\Gamma(\bar q)$ 
that correct for the use of $\bar q$ in the two-body \lamb{3}{H}$\to \pi + 
{^{3}Z}$ rate expressions~\cite{Con92}. We note that applying the $\Delta I 
=\frac{1}{2}$ rule to the isospin $I=0$ decaying \lamb{3}{H}$_{\rm g.s.}$, 
here too as in the free $\Lambda$ decay, the ratio of $\pi^-$:$\pi^0$ 
decay rates is approximately 2:1. The quantity $\eta(\bar q)$ in 
Eq.~(\ref{eq:L3Hg.s.}) is an exchange integral ensuring that the summation 
on final nuclear states is limited to totally antisymmetric states: 
\begin{equation} 
\eta (q) = \int{\chi(\vec{r}_{\Lambda};\vec{r}_{N2},\vec{r}_{N3})
\exp[i\vec{q}\cdot(\vec{r}_{\Lambda}-\vec{r}_{N2})]
\chi^\ast(\vec{r}_{N2};\vec{r}_{\Lambda},\vec{r}_{N3})
{\rm d}^3\vec{r}_{\Lambda}{\rm d}^3\vec{r}_{N2}{\rm d}^3\vec{r}_{N3}}. 
\label{eq:eta1} 
\end{equation} 
Here $\chi(\vec{r}_{\Lambda};\vec{r}_{N2},\vec{r}_{N3})$ is the real 
normalized spatial wavefunction of \lamb{3}{H}, symmetric in the nucleon 
coordinates 2 and 3. This wavefunction, in abbreviated notation $\chi(1;2,3)$, 
is associated with a single spin-isospin term which is antisymmetric in the 
nucleon labels, such that $s_{\Lambda}=\frac{1}{2}$ couples to $\vec{s}_1 + 
\vec{s}_2 = 1$ to give $S_{\rm tot}=\frac{1}{2}$ for the ground state 
and $S_{\rm tot}=\frac{3}{2}$ for the spin-flip excited state (if bound), 
and $t_{\Lambda}=0$ couples trivially with $\vec{t}_1 + \vec{t}_2 = 0$. 
Eq.~(\ref{eq:L3Hg.s.}) already accounts for this spin-isospin algebra in 
\lamb{3}{H}. For completeness we also list the total decay rate expression 
for \lamb{3}{H} if its g.s. spin-parity were $J^P={\frac{3}{2}}^+$: 
\begin{equation} 
\Gamma_{_\Lambda^3{\rm H}}^{J=3/2}=\frac{\bar q}{1+\omega_{\pi}(\bar q)/E_{3N}
(\bar q)}[|s_{\pi}|^2(1-\eta(\bar q))+|p_{\pi}|^2(\frac{\bar q}{q_{\Lambda}})^2
(1-\frac{1}{3}\eta(\bar q))]. 
\label{eq:L3Hexc.} 
\end{equation} 
Since $0<\eta(q)<1$, the dominant $s$-wave term here is weaker than in the 
free $\Lambda$ decay, Eq.~(\ref{eq:GammaLam}), implying that the \lamb{3}{H} 
lifetime would have been longer than the free $\Lambda$ lifetime, had its 
g.s. spin-parity been ${\frac{3}{2}}^+$.

\subsection{\lamb{3}{n}} 
\label{subsec:L3n} 

For \lamb{3}{n}($I=1,J^P={\frac{1}{2}}^+$) weak decay, it is necessary to 
distinguish between decays induced by $\Lambda\to p+\pi^-$ and those induced 
by $\Lambda\to n+\pi^0$. In the first case the spectator neutrons are `frozen' 
to their $s$ shell in both initial and final state, without having to recouple 
spins or consider exchange integrals for the final proton. This means that the 
\lamb{3}{n}~$\to (pnn) + \pi^-$ weak decay rate will be given in the closure 
approximation essentially by the $\Lambda\to p+\pi^-$ free-space weak-decay 
rate. In the other case of $\Lambda\to n+\pi^0$ induced decays, production 
of a low-momentum neutron is suppressed by the Pauli principle on account 
of the two neutrons already there in the initial \lamb{3}{n} state. To a good 
approximation this \lamb{3}{n} weak decay branch may be disregarded. Our best 
estimate for the \lamb{3}{n} weak decay rate is then given by 
\begin{equation} 
\Gamma_{_\Lambda^3{\rm n}}^{J=1/2}\approx\frac{\bar q}{1+\omega_{\pi}(\bar q)
/E_{3N}(\bar q)}\,0.641\,\left(|s_{\pi}|^2+|p_{\pi}|^2(\frac{\bar q}
{q_{\Lambda}})^2\right), 
\label{eq:L3n1} 
\end{equation} 
where the coefficient 0.641 is the free-space $\Lambda\to p+\pi^-$ fraction 
of the total $\Lambda\to N+\pi$ weak decay rate. Evaluating the ratio 
$\Gamma$(\lamb{3}{n})/$\Gamma_\Lambda$ for the choice ${\bar q}=q_\Lambda$ 
one obtains 
\begin{equation} 
\Gamma({_\Lambda^3}{\rm n})/\Gamma_\Lambda \approx 1.114 \times 0.641 = 0.714, 
\label{eq:L3n2} 
\end{equation}  
where the factor 1.114 follows from the difference between $E_{3N}(q_\Lambda)$ 
and $E_{N}(q_\Lambda)$ in the recoil phase-space factors. Our predicted 
\lamb{3}{n} lifetime is then 
\begin{equation} 
\tau({_\Lambda^3}{\rm n}) \approx 368~{\rm ps,} 
\label{eq:L3n3} 
\end{equation} 
but likely not shorter than 350~ps upon assigning 5\% contribution from the 
$\pi^0$ decay branch. This lifetime is way longer than the 181${^{+30}_{-24}}
\pm$25~ps or 190${^{+47}_{-35}}\pm$36~ps lifetimes deduced from the $nd\pi^-$ 
and $t\pi^-$ alleged decay modes of \lamb{3}{n}~\cite{Rap13,Saito16}. Note 
that adding a potentially unobserved proton could perhaps reconcile these 
deduced lifetimes with $\tau$(\lamb{4}{H})=194$^{+24}_{-26}$~ps~\cite{Outa98}.

\section{\lamb{3}{H} lifetime calculation disregarding pion FSI}
\label{sec:noFSI} 

A lesson gained from Ref.~\cite{Con92} is that as long as the binding energy 
of \lamb{3}{H} is reproduced, the lifetime calculation is rather insensitive 
to the fine details of the particular $\Lambda N$ interaction model chosen. 
The main uncertainty in lifetime calculations arises in fact from the 
imprecisely known value of $B_{\Lambda}$(\lamb{3}{H}). Therefore, to have 
as simple input as possible to a three-body description of the weakly bound 
\lamb{3}{H} we constructed baryon-baryon $s$-wave separable interactions 
of Yamaguchi forms by fitting to the corresponding low-energy scattering 
parameters. In particular, the binding energy of the deuteron, limited to 
a $^3S_1$ channel, is reproduced. For the $\Lambda N$ interaction we followed 
Ref.~\cite{GG14} by choosing values of scattering lengths and effective ranges 
close to those used in Nijmegen models. Scaling slightly the $\Lambda N$ 
$^1S_0$ interaction we covered, by solving three-body Faddeev equations, 
a range of $\Lambda$ separation energy $B_\Lambda$ values to account for the 
given uncertainty in the experimental value $B_\Lambda$(\lamb{3}{H})=0.13$
\pm$0.05~MeV. In these calculations the Faddeev integral equations were solved 
using a momentum-space Gauss mesh of 32 points, with half the integration 
points satisfying $q<1\,$fm$^{-1}$, thereby taking good care of the small $q$ 
(large $r$) region which is of utmost importance for the diffuse \lamb{3}{H}. 
The results prove numerically stable already upon using 20 Gauss mesh points. 

To calculate the total decay rate, Eq.~(\ref{eq:L3Hg.s.}), we evaluated the 
exchange integral $\eta (q)$ of Eq.~(\ref{eq:eta1}) by using a wavefunction 
$\Psi(\vec{p}_1,\vec{q}_1)$, derived by solving the appropriate Faddeev 
equations in momentum space in terms of two Jacobi coordinates, say 
$\vec{p}_1$ for the relative coordinate of the two nucleons and $\vec{q}_1$ 
for that of the $\Lambda$ with respect to the center of mass of the nucleons; 
for details see Eqs.~(10--12) and related text in our arXiv:1811.03842v1 
[nucl-th] version. Eq.~(\ref{eq:eta1}) is thereby transformed to 
\begin{equation} 
\eta (q) = \int{\Psi(\vec{p}_1,\vec{q}_1)\,\Psi^\ast(\,\frac{1}{2}\vec{p}_1 + 
\frac{3}{4}\vec{q}_1-\frac{1}{2}\vec{q},\,\,\vec{p}_1-\frac{1}{2}\vec{q}_1
+\vec{q}\,)  
\,{\rm d}^3\vec{p}_1\,{\rm d}^3\vec{q}_1}. 
\label{eq:eta3} 
\end{equation} 

To evaluate this form of the exchange integral $\eta (q)$ one needs to express 
the Faddeev three-body wavefunction $\Psi$ as a function of the two variables 
$\vec{p}_1$ and $\vec{q}_1$. This requires careful attention since the Faddeev 
decomposition $T=T_1+T_2+T_3$ of the total $T$ matrix into three partial $T_j$ 
matrices coupled to each other by the Faddeev equations $T_j=t_j(1+G_0\sum_{
k\neq j}T_k)$ implies a similar decomposition of the bound-state wavefunction 
$\Psi$ into three components 
\begin{equation} 
\Psi = G_0 T \phi = G_0 (T_1+T_2+T_3) \phi \equiv 
\Psi_1 + \Psi_2 + \Psi_3,  
\label{eq:Psi} 
\end{equation} 
where $G_0$ is the three-body free Green's function and $\phi$ is a three-body 
plane wave. The natural momentum variables for each $\Psi_j$ component are 
$\vec{p}_j$ and $\vec{q}_j$, so we need to switch in $\Psi_2$ and $\Psi_3$ 
from their respective momentum bases to $\vec{p}_1$ and $\vec{q}_1$. 
This naturally involves integration on the angle between $\vec{p}_j$ 
and $\vec{q}_j$, $j\neq 1$, so that the redressed $\Psi_2$ and $\Psi_3$ 
necessarily develop $\ell > 0$ partial waves in addition to their dominant 
$s$-wave component. We have omitted such unwanted $\ell \neq 0$ partial waves. 
The error incurred in this approximation may be estimated by evaluating the 
normalization integral of $\Psi$ in two ways, first with each component 
$\Psi_j$ in its natural coupling scheme, and then with all three components 
expressed in terms of the $\vec{p}_1,\vec{q}_1$ variables, thus giving rise 
to a normalization integral smaller by about 2.5\%. 

The \lamb{3}{H} exchange integral $\eta (q)$ of Eq.~(\ref{eq:eta3}) was 
evaluated numerically for two values of the closure momentum $\bar q$ 
discussed in Ref.~\cite{RD66} and for several values of $B_\Lambda
$(\lamb{3}{H}) suggested by its experimental uncertainty. Our results are 
listed in Table~\ref{tab:eta}, compared with those of Congleton~\cite{Con92} 
who considered the same range of $B_\Lambda$(\lamb{3}{H}) values.   

\begin{table}[htb]
\begin{center}
\caption{Values of the \lamb{3}{H} exchange integral $\eta(\bar q)$ in the 
present $\Lambda pn$ Faddeev approach, evaluated for two representative values 
of the closure momentum $\bar q$, and the value calculated by Congleton for 
${\bar q}=96$~MeV/c using a $\Lambda d$ cluster description of \lamb{3}{H} 
\cite{Con92}. The listed uncertainties reflect the uncertainty in the value of  
$B_{\Lambda}$(\lamb{3}{H})=0.13$\pm$0.05~MeV.} 
\begin{tabular}{ccc} 
\hline 
Model & $\bar q$ (MeV/c) & $\eta(\bar q)$ \\ 
\hline 
$\Lambda d$ cluster~\cite{Con92} &  96 & 0.212$\pm$0.011 \\ 
$\Lambda pn$ Faddeev [present]   &  96 & 0.146$\pm$0.021 \\ 
$\Lambda pn$ Faddeev [present]   & 104 & 0.130$\pm$0.021 \\ 
\hline 
\end{tabular} 
\label{tab:eta}
\end{center}
\end{table} 

As argued by RD~\cite{RD66}, and followed up by Congleton~\cite{Con92}, 
the appropriate choice for the \lamb{3}{H} pionic decay closure momentum 
$\bar q$ is the empirical peak value $\bar q = 96$~MeV/c in the $\pi^-$ weak 
decay continuum spectrum. To study the sensitivity of $\eta (\bar q)$ to 
a small departure from this accepted value of $\bar q$, we also evaluated 
$\eta (\bar q)$ for $\bar q = 104$~MeV/c, a value a bit larger than that 
for the free $\Lambda$ decay which was used in calculations that preceded RD. 
The variation of $\eta (\bar q)$ with $\bar q$ over the momentum interval 
studied is quite weak. For ${\bar q} = 96$~MeV/c, our calculated value of 
$\eta (\bar q)$ is about 70\% of Congleton's value. This apparent discrepancy 
can be shown to arise from his use of a $\Lambda d$ cluster model for 
\lamb{3}{H}: specifically by (i) limiting the full Faddeev wavefunction 
$\Psi$, Eq.~(\ref{eq:Psi}), to $\Psi_1$ which is the component most natural 
to represent a $\Lambda d$ cluster, and (ii) suppressing then in the 
three-body free Green's function $G_0$ the dependence on the $\Lambda$ 
momentum, we obtain a value of $\eta (\bar q = 96$~MeV/c)=0.238$\pm$0.038, 
in good agreement with the value listed for '$\Lambda d$ cluster' in 
Table~\ref{tab:eta}. 

Using $\eta ({\bar q}=96$~MeV/c)=0.146$\pm$0.021 from Table~\ref{tab:eta} in 
Eq.~(\ref{eq:L3Hg.s.}) for \lamb{3}{H}, and noting Eq.~(\ref{eq:GammaLam}) 
for the free $\Lambda$ decay, we obtain the following \lamb{3}{H} mesonic 
decay rate: 
\begin{equation} 
\Gamma_{_\Lambda^3{\rm H}}^{J=1/2}/
\Gamma_\Lambda = 1.09\pm 0.01,~~~~{\rm or}~~\tau({_\Lambda^3{\rm H}})/
\tau_\Lambda = 0.92\pm 0.01. 
\label{eq:final1} 
\end{equation} 
Adding a $\approx$1.7\% nonmesonic weak decay branch~\cite{Golak97}, 
our final result is 
\begin{equation} 
\tau({_\Lambda^3{\rm H}})/\tau_\Lambda = 0.90\pm 0.01.
\label{eq:final2} 
\end{equation}

\section{Pion FSI effects} 
\label{sec:pion} 

The pion emitted in the \lamb{3}{H} decay is dominantly $s$-wave pion, see 
Eq.~(\ref{eq:GammaLam}). Optical model fits of measured $1s$ pionic atom level 
shifts and widths across the periodic table~\cite{BFG97,FG07} suggest that the 
underlying $\pi N$ $s$-wave interaction term in nuclei at low energy is weakly 
repulsive and that the attractive $\pi N$ $p$-wave term has negligible effect 
on $1s$ pionic states. The corresponding $s$-wave induced $\pi^-$ nuclear 
optical potential is given by{\footnote{We disregard here a weak two-nucleon 
absorptive term in order to retain the few percent unobserved decay branch 
\lamb{3}{H}$\,\to pnn$.}} 
\begin{equation} 
V_{\rm opt}^{\pi^-}=-\frac{4\pi}{2\mu_{\pi N}}
\left(b_0[\rho_n(r)+\rho_p(r)]+b_1[\rho_n(r)-\rho_p(r)]\right), 
\label{eq:Vopt} 
\end{equation} 
in terms of fitted real $\pi N$ scattering lengths: isoscalar $b_0=-0.0325$~fm 
and isovector $b_1=-0.126$~fm~\cite{FG14}. With these negative signs one gets 
$\pi^-$ repulsion in the majority of stable nuclei, those with $N\geq Z$. 
However, in the few $Z>N$ available nuclear targets like $^1$H and $^3$He 
this repulsion is reversed into attraction owing to the isovector term 
flipping sign under $Z\leftrightarrow N$. This is confirmed by the attractive 
$1s$ level shifts observed in the $\pi^-\,{^1}$H and $\pi^-\,{^3}$He 
atoms~\cite{Gotta04}. One therefore expects attractive FSI in the $^3$He+$
\pi^-$ decay channel of \lamb{3}{H}, and repulsive FSI in the $^3$H+$\pi^0$ 
decay channel where the $\pi N$ isovector term associated with $b_1$ 
vanishes while the weakly repulsive isoscalar term associated with $b_0$ 
remains in effect. Altogether we have verified that the sum of these two 
FSI contributions to the \lamb{3}{H} decay rate is nearly zero. Pion FSI 
in the context of \lamb{3}{H} decay has been considered elsewhere only 
by RD~\cite{RD66}, who indeed found it weakly repulsive and lowering 
the \lamb{3}{H} decay rate by 1.3\%, and in Ref.~\cite{Kelkar97} where 
the attraction in the $^3$He+$\pi^-$ decay channel was overlooked. 

The preceding argumentation on the role of pion FSI in the \lamb{3}{H} 
decay is incomplete, if not misleading. The $\Delta I=\half$ rule in the 
$\Lambda\to N\pi$ weak decay implies that the $3N+\pi$ final states are good 
($I=\half,\,I_z=-\half$) isospin states which are coherent combinations of 
$ppn\pi^-$ and $pnn\pi^0$ configurations. For $I=\half$, and in the Born 
approximation applied to the optical potential Eq.~(\ref{eq:Vopt}), the 
pion-nuclear scattering length is given by $3b_0-2b_1$ which is considerably 
more {\it attractive} than the scattering length $3b_0-b_1$ valid for 
$^3$He+$\pi^-$ alone. The difference between these two expressions arises 
from charge exchange transitions between the nearly degenerate charge states 
of ($I=\half,I_z=-\half$) good-isospin states, such as between $^3$He+$\pi^-$ 
and $^3$H+$\pi^0$. To estimate the effect of pion FSI on the \lamb{3}{H} 
lifetime we consider \lamb{3}{H} decays to good-isospin states made of the 
corresponding two nuclear charge states. 

In the distorted-wave (DW) approximation, the \lamb{3}{H} decay amplitude is 
given by a form factor
\begin{equation} 
F_{\rm DW}(q)=\int{\Phi^\ast_{3N}({\vec r},{\vec \rho})\,\tilde{j_0}(qr_N)\,
\Phi_{_\Lambda^3{\rm H}}({\vec r},{\vec \rho})\,{\rm d}^3r\,{\rm d}^3\rho}, 
\label{eq:F(q)} 
\end{equation} 
where $\tilde{j_0}$ is a pion DW evolving via FSI from a pion plane-wave (PW) 
spherical Bessel function $j_0$. The vectors $\vec r$ and $\vec \rho$ are 
Jacobi coordinates: $\vec r$ stands for the $\Lambda\to N$ `active' 
baryon relative to the cm of the spectator nucleons, and ${\vec\rho}$ denotes 
the relative coordinate of the spectator nucleons. In Eq.~(\ref{eq:F(q)}), 
${\vec r_N}=\frac{2}{3}{\vec r}$ stands for the coordinate of the `active' 
baryon with respect to the cm of the $3N$ final system. The $\Phi_\alpha$ 
are properly normalized $L=0$ initial and final $A=3$ wavefunctions. For 
this first evaluation we approximated each $\Phi_\alpha({\vec r},{\vec\rho})$ 
by a product form $\psi_\alpha(r)\phi_\alpha(\rho)$, with single-baryon 
bound-state wavefunctions $\psi_\alpha(r)$ given by 
\begin{equation} 
r\psi_\alpha(r)\sim \exp(-\kappa_\alpha r) - \exp(-\beta_\alpha r)
\label{eq:psi} 
\end{equation} 
as generated by Yukawa separable potentials. The choice of the inverse range 
parameters $\kappa_\alpha$ and $\beta_\alpha$ is discussed below following 
Table~\ref{tab:DW}. With a product form of $\Phi_\alpha({\vec r},{\vec\rho})$, 
the form factor (\ref{eq:F(q)}) reduces to 
\begin{equation} 
F_{\rm DW}(q)=\gamma_{\rm d}\,\int{\psi^\ast_N(r)\,\tilde{j_0}(qr_N)\,
\psi_\Lambda(r)\,{\rm d}^3r},  
\label{eq:F(q)red} 
\end{equation} 
where 
\begin{equation}
\gamma_{\rm d}=\int{\phi^\ast_{3N}(\rho)\phi_{_\Lambda^3{\rm H}}
(\rho)\,{\rm d}^3\rho} 
\label{eq:gamma} 
\end{equation} 
is the overlap integral of the two $\phi$s and is the same for both PW 
and DW pions. For this reason, the choice of the `deuteron' wavefunctions 
$\phi_\alpha(\rho)$ is not discussed further here. For the pion DW 
$\tilde{j_0}$ we used a continuum wavefunction, also generated from 
a separable Yukawa potential: 
\begin{equation} 
\tilde{j_0}(qr_N)=j_0(qr_N)+\frac{f(q)}{r_N}\left(\exp(iqr_N)-\exp(-\beta_\pi 
r_N)\right), 
\label{eq:DW} 
\end{equation} 
where $f(q)$ is a $\pi$-nuclear $s$-wave scattering amplitude derived from 
SAID~\cite{SAID06} $\pi N$ partial-wave amplitudes, with values also listed 
in Table~\ref{tab:DW}. 

\begin{table}[htb]
\begin{center}
\caption{Parameters of $N$, $\Lambda$ and $\pi$ wavefunctions (\ref{eq:psi}) 
and (\ref{eq:DW}) used to evaluate the pion DW amplitude $F_{\rm DW}(q)$, 
Eq.~(\ref{eq:F(q)red}), for two-body and three-body \lamb{3}{H} decays, see 
text. Values of $\kappa$ and $\beta$ are in fm$^{-1}$ units, $f(q)$ in fm.} 
\begin{tabular}{cccccccc} 
\hline 
decay to & $\kappa_\Lambda$ & $\kappa_N$ & $\beta_\Lambda$ & $\beta_N$ & 
$\beta_\pi$ & $f(q)$ & $|F_{\rm DW}/F_{\rm PW}|^2$ \\ 
\hline 
$\pi$+$^3$Z & 0.068 & 0.420 & 1.2 & 1.2 & 0.806 & 0.180+i0.048 & 1.097 \\ 
$\pi$+$N$+$d$ & 0.068 & 0.068 & 1.2 & 1.2 & 1.626 & 0.225+i0.022 & 1.119 \\ 
\hline 
\end{tabular} 
\label{tab:DW} 
\end{center} 
\end{table}

\subsection{Two-body decay modes} 
\label{subsec:2-body} 

The input parameters for the evaluation of pion FSI effects on the two-body 
decay modes \lamb{3}{H}$\,\to$$\pi+^3$Z are given in the first row of 
Table~\ref{tab:DW}. The listed values of each wave number $\kappa_\alpha$, 
Eq.~(\ref{eq:psi}), follow from the separation energy of the corresponding 
baryon $\alpha$ with respect to the deuteron in the initial hypernucleus 
and final nucleus. The choice of $\beta_N$ corresponds to r.m.s. radius 
$<r^2>^{1/2}_{\psi_N}=2.631$~fm which for a spatially symmetric $^3$He 
wavefunction reproduces its matter r.m.s. radius $<r_N^2>^{1/2}_{^3{\rm He}}
=1.754$~fm. The choice of $\beta_\Lambda$ hardly matters in reproducing 
the expectedly large r.m.s. radius of the loosely bound \lamb{3}{H}, 
which is 10.4~fm for $\Lambda$ relative to $d$ using the asymptotic term 
$\exp(-\kappa_\Lambda r)$ of $\psi_\Lambda$. For convenience we chose 
$\beta_\Lambda=\beta_N$. As for the pion DW, Eq.~(\ref{eq:DW}), $\beta_\pi$ 
was determined by equating the r.m.s. radius of a Yukawa form factor $g(r_N)=
\exp(-\beta_\pi)/r_N$, for a separable pion potential $g(r_N)g(r_N')$, to 
the $^3$He matter r.m.s. radius: $\beta_\pi =\sqrt{2}/1.754$~fm$^{-1}$. 
Finally, the value of the pion-nuclear scattering amplitude $f(q)=3b_0-2b_1$ 
listed in the table was derived using values of $b_0$ and $b_1$ taken from 
SAID~\cite{SAID06} at 31~MeV pion kinetic energy, compared to 0.155~fm for 
the threshold values listed following Eq.~(\ref{eq:Vopt}). The variation of 
both $b_0$ and $b_1$ with energy in the SAID analysis is rather weak. In the 
actual calculation we dropped the small imaginary part of $f(q)$ so as to 
account also for the few percent pion absorption contribution \lamb{3}{H}$\,
\to pnn$ to the \lamb{3}{H} lifetime. 

Evaluating the form factors $F_{\rm DW}(q)$ and $F_{\rm PW}(q)$, where the 
latter is obtained from the former by reducing $\tilde{j_0}$ to $j_0$, we 
find a pion FSI enhancement factor $|F_{\rm DW}(q)/F_{\rm PW}(q)|^2 = 1.097$ 
for the \lamb{3}{H} decay rate to the two-body final states $^3$He+$\pi^-$ 
and $^3$H+$\pi^0$. With this enhancement, these two-body decay modes take 
roughly 35.7\% of the total decay rate in the present $s$-wave calculation, 
in agreement with the branching ratio 0.35$\pm$0.04 extracted from $\pi^-$ 
decays in helium bubble chamber measurements~\cite{KSWB73}.

\subsection{Three-body decay modes} 
\label{subsec:3-body} 

Evaluating the pion FSI in the three-body \lamb{3}{H} decays to $p+d+\pi^-$ 
and $n+d+\pi^0$ final states is more involved than done above for the two-body 
decay modes because the nuclear bound state wavefunction $\psi_N$ has to 
be replaced by continuum nucleon wavefunctions. At this stage we report on 
a rough approximation that still uses $\psi_N$ of the form (\ref{eq:psi}), 
but for a loosely bound nucleon on its way to become unbound. We chose for 
convenience $\kappa_N=\kappa_\Lambda$, as listed in the second row of 
Table~\ref{tab:DW}, having verified that the resulting enhancement factor 
$|F_{\rm DW}(q)/F_{\rm PW}(q)|^2$ hardly changes near threshold, although 
each of the separate form factors does. For the pion distorted wave we assumed 
it is dominated by FSI with the outgoing nucleon, since the low-energy $\pi d$ 
interaction is much weaker than its $\pi N$ counterpart. The pion-nucleon 
inverse-range parameter $\beta_\pi$ listed in the table corresponds to 
a nucleon r.m.s. radius of 0.87~fm, and its scattering amplitude listed in the 
table corresponds to a combination $f(q)=b_0-2b_1$ using values of $b_0$ and 
$b_1$ again derived from SAID~\cite{SAID06}. This rough estimate gives a 1.119 
enhancement factor for the three-body decay modes. 

Given the \lamb{3}{H} decay rate enhancement factors $|F_{\rm DW}(q)/
F_{\rm PW}(q)|^2$ listed in Table~\ref{tab:DW} for two- and three-body final 
states, which according to Ref.~\cite{Fad98} saturate more than 98\% of the 
pionic decay modes, the total pion FSI enhancement factor is (11$\pm$2)\%, 
where the 2\% uncertainty was estimated by varying the $\Lambda$ and $N$ 
inverse-range parameters listed in the table within reasonable limits. 
Varying the pion inverse-range parameter $\beta_\pi$ introduces larger 
uncertainty, and this will have to be studied more quantitatively in future 
four-body calculations.

\section{Conclusion} 
\label{sec:concl} 

In this work we evaluated the mesonic decay rate of \lamb{3}{H} by considering 
the closure-approximation decay-rate expression Eq.~(\ref{eq:L3Hg.s.}). 
The \lamb{3}{H} exchange integral $\eta (\bar q)$, Eq.~(\ref{eq:eta1}), 
which provides input to Eq.~(\ref{eq:L3Hg.s.}) was calculated within 
a fully three-body Faddeev equations model of \lamb{3}{H}$_{\rm g.s.}$. For 
${\bar q}=96$~MeV/c, as suggested by RD~\cite{RD66}, our calculated value of 
$\eta (\bar q)$ listed in Table~\ref{tab:eta} leads to a (9$\pm$1)\% increase 
of the \lamb{3}{H} mesonic decay rate over the free $\Lambda$ decay rate. This 
result supersedes Congleton's result~\cite{Con92} of 12\% increase based on a 
$\Lambda d$ cluster model of \lamb{3}{H} which gave a value of $\eta(\bar q)$ 
about 50\% higher than our Faddeev equations model value. Adding a 1.7\% 
nonmesonic decay rate contribution~\cite{Golak97} we get a (10$\pm$1)\% 
decrease of the \lamb{3}{H} lifetime with respect to the free $\Lambda$ 
lifetime, a stronger decrease than the 6\% derived in the more microscopically 
oriented Faddeev calculation by Kamada et al.~\cite{Fad98}. Neither of these 
calculations is sufficient on its own to resolve the \lamb{3}{H} lifetime 
puzzle. 

Perhaps more importantly, we considered the pion FSI effect on the \lamb{3}{H} 
lifetime. This effect goes beyond any pion FSI effect already present in 
the empirical $\Lambda\to N +\pi$ decay parameters $s_{\pi}$ and $p_{\pi}$. 
Simple arguments rooted in low-energy pion-nucleon and pion-nucleus 
phenomenology were shown to imply an attractive FSI, an observation that 
escaped the attention of all previous works. This attractive pion FSI 
was evaluated here semi-quantitatively and found to shorten further 
the \lamb{3}{H} lifetime down to (81$\pm$2)\% of $\tau_\Lambda$. Further, 
although little reduction of $\tau$(\lamb{3}{H}) could arise from attractive 
$p$-wave pion FSI contributions. More involved calculations going beyond 
three-body calculations are required to verify the overall substantial 
reduction owing to the pion FSI in this $A=3$ system. 

Last, as a by-product of our formulation of the $A=3$ hypernuclear lifetime, 
we showed in simple terms that the lifetime of \lamb{3}{n}, if bound, 
is considerably longer than $\tau_\Lambda$, in disagreement with the 
shorter lifetime with respect to $\tau_\Lambda$ extracted from the HypHI 
events assigned to this hypernucleus. Pion FSI should be repulsive in this 
case, aggravating this disagreement by increasing further the \lamb{3}{n} 
lifetime by a few more percents.

\section*{Acknowledgements} 

A.G. acknowledges useful discussions with, and advice from Eli Friedman 
on pion FSI. H.G. acknowledges support by COFAA-IPN (M\'exico).


\end{document}